\documentclass{article}
\usepackage[cp1251]{inputenc}
\usepackage[russian]{babel}

\begin{document}
\title{GENERALIZED-ANALYTICAL FUNCTIONS OF \\
 POLY-NUMBER VARIABLE}

\author{G. I. Garasko \\ Electrotechnical institute of Russia,\\
 gri9z@mail.ru }

\maketitle
\begin{abstract}

    We introduce the notion of the generalized-analytical function of
the poly-number variable, which is a non-trivial generalization of
the notion of analytical function of the complex variable and,
therefore,  may turn out to be fundamental in theoretical physical
constructions. As an example we consider in detail  the
associative-commutative hypercomplex numbers $H_4$ and an
interesting class of corresponding functions.
\end{abstract}

\section{Introduction}

Let $M_n$  be an $n$-dimensional elementary manifold and $P_n$ denote the
system of  $n$-dimensional associative-commutative hypercomplex numbers
(poly-numbers, $n$-numbers), and a one-to-one correspondence between the sets
be assigned. Under these conditions, we choose in $P_n$  the basis
\begin{equation}\label{1}
e_1,e_2,...,e_n; \quad e_ie_j=p^k_{ij}e_k,
\end{equation}
\begin{equation}\label{2}
 X=x^1\cdot e_1+x^2\cdot e_2+\cdots+x^ne_n\in P_n,
\end{equation}
where $e_1,e_ 2,...,e_n$ - symbolic elements, $ p^k_{ij}$ stand
for characteristic  real numbers, and $x^1,x^2,...,x^n$  -  real
coordinates with respect  to  the basis ($ e_1\equiv
1,e_2,\dots,e_n$). Obviously, the numbers $x^1,x^2,...,x^n$
 can be used not only as the coordinates in $P_n$,
  but also as coordinates in the manifold $M_n$ , so that $(x^1,x^2,...,x^n)\in M_n$.
  Though in $M_n$ we can go over to any other curvilinear reference frame,
  the reference frame $\{x^i\}$, as being built by the help the basis
   of poly-numbers and a fixed one-to-one correspondence
   $M_n\leftrightarrow P_n$,
    ought to be considered preferable
    (as well as any other reference frame
     connected with this by non-degenerate linear transformation).
     The poly--number algebraic operations induce the same operations
     in the elementary manifold (formally) and
      in the tangent space at any point of the manifold (informally).
    Accordingly, the tangent spaces to $M_n$  are isomorphic to  $P_n$.

       The function
\begin{equation}\label{3}
F(X) := f^1(x^1,...,x^n)e_1+...+f^n(x^1,...,x^n)e_n
\end{equation}

of the poly-number variable, where $f^i$ are sufficiently smooth
functions of $n$ real variables, will be considered to be a vector
(contravariant) field in $M_n$. Hence, apart from addition and
multiplication by number,  any operation of multiplication of
vector fields
\begin{equation}\label{4}
 f_{(3)}^k=f_{(1)}^i\cdot f_{(2)}^j\cdot p^k_{ij}
\end{equation}

can also defined in $M_n$. It is useful but not obligatory to
consider the space $M_n$ to be the  main (``the examined") object
and the space $P_n$ to be a sort of an instrument with the help of
which the space $M_n$ is ``examined". In the general case the
parallel transportation of a vector in the space $P_n$ does not
correspond to the ``parallel transportation"\, of the same vector
in the space $M_n$, so that for a due definition of absolute
differential (or the covariant derivative) we are to have the
connection objects or the quantities which may replace them. If we
avoid introducing the pair $\{ M_n, P_n\}$,  restricting the
treatment only to associative-commutative hypercomplex numbers,
then it is natural to introduce the definitions
\begin{equation}\label{5}
dX:= dx^i\cdot e_i
\end{equation}
and
\begin{equation}\label{6}
dF(X) :=F(X+dX)-F(X)=\frac{\partial f^i}{\partial x^k}\cdot
e_i\cdot dx^k.
\end{equation}
 The function $F(X)$  of poly-number variable  $X$ is called \textit{analytical},
 if such a function  $F'(X)$ exists that
\begin{equation}\label{7}
dF(X)=F'(X)\cdot dX,
\end{equation}
where the multiplication in the right-hand part means the
poly-number operation. From (7) it follows that
\begin{equation}\label{8}
\frac{\partial f^i}{\partial x^k}=p^i_{kj}\cdot f'^j.
\end{equation}
Since with respect to the basis $e_i$  with the components $e_1=1$
the equalities
\begin{equation}\label{9}
p^i_{1j}=\delta^i_j
\end{equation}
   hold,  we have
\begin{equation}\label{10}
f'^i=\frac{\partial f^i}{\partial x^1}.
\end{equation}
Inserting (10) in (8) yields the Cauchy-Riemann relations
\begin{equation}\label{11}
 \frac{\partial f^i}{\partial x^k}-p^i_{kj}\cdot
 \frac{\partial f^j}{\partial x^1}=0
\end{equation}
for the functions under study. The number $n(n-1)$ of these
relations is growing quicker that the number $n$ of components of
analytical function. This leads to the \textit{functional
restriction} of the set of such functions at $n>2$. The present
work is just attempting to elaborate  a non-trivial extension of
the notion of analytical function of poly-number variable subject
to the condition that number of the Cauchy-Riemann-type conditions
does nor exceed the number of unknown function--components. The
first step in this direction has been made above when introducing
the pair $\{M_n, P_n\}$. Therefore it seems natural to replace the
differential 6) by means of the absolute differential
\begin{equation}\label{12}
DF(X) :=\nabla_kf^i\cdot e_i\cdot dx^k,
\end{equation}
where
\begin{equation}\label{13}
\nabla_kf^i := \frac{\partial f^j}{\partial x^k} + \Gamma^i_{kj}
\cdot f^j
\end{equation}
is  the covariant derivative, and $\Gamma^i_{kj}$ means "the
connection coefficients". Instead of the formulas (8) and  (10) we
get
\begin{equation}\label{14}
\nabla_kf^i =  p^i_{kj}\cdot f'^j
\end{equation}
and
\begin{equation}\label{15}
f'^i=\nabla_1\cdot f^i,
\end{equation}
and the Cauchy-Riemann conditions take on the form
\begin{equation}\label{16}
\nabla_kf^i - p^i_{kj}\cdot \nabla_1f^j  =0.
\end{equation}
Of course, "the connection objects"\, $\Gamma^i_{kj}$ in the
formula (13) are not obligatory to be uniform for all the set of
functions obeying the conditions (16).

\section{Definitions and basic implications}

Let us call the function $F(X)$  \textit{generalized-analytical},
if such a function $F'(X)$ exists that
\begin{equation}\label{17}
\tilde{D}F(X)=F'(X)\cdot dX,
\end{equation}
where
\begin{equation}\label{18}
\tilde{D}F(X) \equiv \tilde{\nabla}_kf^i\cdot e_i\cdot dx^k
\end{equation}
and the definition
\begin{equation}\label{19}
_kf^i :=
 \frac{\partial f^i}{\partial x^k}+\gamma^i_k
\end{equation}
has been used. It is assumed that under the transition from one
(curvilinear) coordinate system to another coordinate system the
involved objects $ \gamma^i_k $ are transformed according to the
law
\begin{equation}\label{20}
\gamma^{i'}_{k'}= \frac{\partial x^k}{\partial x^{k'}} \cdot
\frac{\partial x^{i'}}{\partial x^i}
 \cdot \gamma^i_k - \frac{\partial x^k}{\partial x^{k'}} \cdot \frac{\partial^2
 x^{i'}}{\partial  x^k \partial x^i} \cdot f^i.
\end{equation}
It will be noted that such a  definition entails that
$\tilde{\nabla}_kf^i$ behaves like a tensor. The quantities
$\gamma^i_k$ will be called the \textit{gamma-objects}. In general
we do not assume the relations
\begin{equation}\label{21}
\gamma^i_k  =\Gamma^i_{kj}\cdot f^j
\end{equation}
with a single "connection object"\, $ \Gamma^i_{kj} $ for
generalized-analytical functions. It would be more precise to say
of the pair $\{f^i,\gamma^i_k\} $, such that the analytical
function of poly-number variable is the pair $ \{ f^i,0\} $, but
this pair transform to the pair $ \{f^{i'},\gamma^{i'}_{k'}\ne
0\}$ under going over from the special coordinate system to
another curvilinear one.

From the definition of  generalized-analytical functions it follows that
\begin{equation}\label{22}
\tilde{\nabla}_kf^i=p^i_{kj}\cdot f'^j
\end{equation}
and
\begin{equation}\label{23}
f^{'i}
 =\tilde{\nabla}_1f^i;
\end{equation}
the respective generalized Cauchy-Riemann relations take on the
form
\begin{equation}\label{24}
\tilde{\nabla}_kf^j-p^i_{kj} \tilde{\nabla}_1f^j=0.
\end{equation}

    The number of unknown functions in the pair $\{f^i,\gamma^i_k\}$  equals
$n+n^2=n(n+1)$,  -- which is more than the number $n(n-1)$ of the
generalized Cauchy-Riemann relations (24). Thus, to use the notion
of  generalized-analytical function in theoretical-physical
constructions it is necessary to additionally establish and
formulate the set of requirements (possibly  one requirement)
which, when used in conjunction with the notion of
generalized-analytical function, would lead unambiguously to
equations of some field of physical meaning. Usually, they are $n$
partial differential equations of second order for
  $n$ independent function-component field.

If $\{f^i_{(1)},\gamma^i_{(1)k}\}$  and
$\{f^i_{(2)},\gamma^i_{(2)k}\}$ - two generalized-analytical
functions, then their arbitrary linear sum with real coefficients
$\alpha,\beta$  is a generalized-analytical function. This ensues
directly from the definition, on using also the formulae (22)-(24)
and (20). Thus, we have
\begin{equation}\label{25}
\alpha \cdot \{ f^i_{(1)},\gamma^i_{(1)k} \} + \beta \cdot
\{f^i_{(2)},\gamma^i_{(2)k}\} = \{ \alpha \cdot f^i_{(1)} + \beta
\cdot f^i_{(2)}, \alpha \cdot \gamma^i_{(1)k} + \beta \cdot
\gamma^i_{(2)k}\}.
\end{equation}

     Now, let us consider the poly-number product of two
generalized-analytical functions $f^i_{(1)}$ and $f^j_{(2)}$:
\begin{equation}\label{26}
f^k_{(3)}= f^i_{(1)}\cdot  f^j_{(2)}\cdot p^k_{ij}
\end{equation}
and try to find the object $\gamma^i_{(3)k}$ such that the pair
$\{f^i_{(3)},\gamma^i_{(3)k}\}$  be generalized-analytical
function. To this end we formally differentiate the left and right
parts of (26) with respect to $x^k$  and use the formula (22),
obtaining
\begin{equation}\label{27} \frac{ \partial
f^i_{(3)}}{\partial x^k} + \gamma^i_{(3)k} =
 p^{i_1}_{kj}p^i_{i_1i_2}f'^j_{(1)}f'^{j_2}_{(2)}
 +
p^{i_2}_{kj}p^i_{i_1i_2}f^{i_1}_{(1)}f'^j_{(2)}.
\end{equation}
Owing to the formula
\begin{equation}\label{28}
p^r_{im}\cdot p^m_{kj}=p^r_{km}\cdot p^m_{ij}
\end{equation}
(which is an implication of the properties of associativity and
commutativity of poly-numbers), we can write
\begin{equation}\label{29}
\frac{ \partial f^i_{(3)}}{\partial x^k} + \gamma^i_{(3)k} =
p^{i}_{kj}p^j_{i_1i_2}(f'^{i_1}_{(1)}f^{i_2}_{(2)} +
f^{i_1}_{(1)}f'^{i_2}_{(2)}),
\end{equation}
where
\begin{equation}\label{30}
\gamma^i_{(3)k}= p^i_{i_1i_2} \cdot
(\gamma^{i_1}_{(1)k}f^{i_2}_{(2)} + f^{i_1}_{(1)}
\gamma^{i_2}_{(2)k}).
\end{equation}
The result (29) can conveniently be represented in terms of the
absolute differential as follows:
\begin{equation}\label{31}
D[F_{(1)}(X)\cdot F_{(2)}(X)]=[DF_{(1)}(X)]\cdot F_{(2)}(X) +
F_{(1)}(X)\cdot [DF_{(2)}(X)]
\end{equation}
or
\begin{equation}\label{32}
D[F_{(1)}(X)\cdot F_{(2)}(X)]= [F'_{(1)}(X)\cdot F_{(2)}(X) +
F_{(1)}(X)\cdot F'_{(2)}(X)]\cdot dX.
\end{equation}
From the last formula we obtain the relation
\begin{equation}\label{33}
[F_{(1)}(X)\cdot F_{(2)}(X)]'= F'_{(1)}(X) \cdot F_{(2)}(X) +
F_{(1)}(X) \cdot F'_{(2)}(X).
\end{equation}

    It remains to clarify whether the transformation law of the
objects $\gamma^i_{(3)k}$ under the transitions to arbitrary
coordinate system is correct. With this aim the formula (30)
should be written in a varied form:
\begin{equation}\label{34}
\gamma^i_{(3)k}=p^i_{i_1i_2}(\gamma^{i_1}_{(1)k}f^{i_2}_{(2)} +
 f^{i_1}_{(1)} \gamma^{i_2}_{(2)k}) +
 (\Gamma^i_{km}p^m_{i_1i_2} - \Gamma^m_{ki_1}p^i_{mi_2} - \Gamma^m_{ki_2}p^i_{i_1m})
 f^{i_1}_{(1)} f^{i_2}_{(2)},
\end{equation}
where  $\Gamma^j_{im}\equiv 0$  with the respect to our special
coordinate system; however, under the transition to an arbitrary
coordinate system the objects $\Gamma^j_{ik}$ transform like
ordinary connection objects and in general
$\Gamma^{j'}_{i'k'}\ne0$ . The condition $\Gamma^j_{ik}\equiv 0$
can also be replaced to apply the more general condition
\begin{equation}\label{35}
\Gamma^i_{km}p^m_{i_1i_2}- \Gamma^m_{ki_1}p^i_{mi_2} -
\Gamma^m_{ki_2}p^i_{i_1m} \equiv 0
\end{equation}
and, moreover, the three coefficients $\Gamma$  in (35) can be
regarded as different. It is possible to restrict ourselves to but
the class of generalized-analytical function obeying the property
\begin{equation}\label{36}
({}^{(1)}\Gamma^i_{km}p^m_{i_1i_2} -
{}^{(2)}\Gamma^m_{ki_1}p^i_{mi_2} -
{}^{(3)}\Gamma^m_{ki_2}p^i_{i_1m})\cdot f^{i_1}_{(1)}
f^{i_2}_{(2)} \equiv 0.
\end{equation}

    Given  the special coordinate system. If one has
$ \Gamma^i_{jk} \equiv {}^{(1)}\Gamma^i_{jk} \equiv
{}^{(2)}\Gamma^i_{jk} \equiv {}^{(3)}\Gamma^i_{jk} \equiv 0, $
 then the tensor  $p^k_{ij}$ is transported
"parallel" \, without any changes in components.

Thus, the poly-product of two generalized-analytical functions of
poly-number variable is again a generalized-analytical function,
and the formula (33) takes place for derivatives if one adopts
that the "connection coefficients"\, associated to the tensor
$p^k_{ij}$ with respect to the special coordinate system vanishes
identically over all three indices. In terms of the pairs
$\{f^i,\gamma^i_k\}$ the poly-product of two
generalized-analytical function can be written as follows:
\begin{equation}\label{37}
\{f^{i_1}_{(1)},\gamma^{i_1}_{(1)}\} \cdot
\{f^{i_2}_{(2)},\gamma^{i_2}_{(2)}\} =
 \{p^i_{i_1i_2} f^{i_1}_{(1)}f^{i_2}_{(2)}, \, p^i_{i_1i_2} \cdot
(\gamma^{i_1}_{(1)k}f^{i_2}_{(2)} +
f^{i_1}_{(1)}\gamma^{i_2}_{(2)k})\}.
\end{equation}
So, the polynomial or the converged series with real or
poly-number coefficients of one or several
 generalized-analytical functions is a  generalized-analytical function.
The ordinary differentiation rules are operative for the respective derivative
(which was denoted my means of the prime $(')$)
of such polynomials and series, whenever
the tensor  $p^k_{ij}$ with
respect to the special coordinate system vanishes identically over all three indices.

 Since in such a theory of generalized-analytical functions of poly-number variable
 (in which the  "connection objects" \, as well as the gamma-objects are different for
 each tensor and, generally speaking, for each index),
 the concept of  "parallel transportation" \, is deprived of the geometrical simplicity
 that is characteristic of the spaces of affine connection,
 the Riemannian and pseudo-Riemannian spaces included.
  This notwithstanding, the concepts of absolute differential and covariant derivative
can readily be extended on the basis of invariance of their form
with respect to any curvilinear coordinate system. The covariant
derivative  $ \tilde{\nabla}_k $  for arbitrary tensor is defined
quite similarly to  the way which is followed to define the
covariant derivative
 $\tilde{\nabla}_k $  in the spaces of affine connection; at the same time, for each tensor
 and probably for each index there exist, in general, their own
 "connection objects" \, or gamma-objects.
 The respective differential is constructed in accordance with the definition
\begin{equation}\label{38}
\tilde{D} := dx^k\cdot \tilde{\nabla}_k.
\end{equation}
Here, the converted indices can not be ignored, for  "connection
coefficients" \,correspond  to them.

  The Cauchy-Riemann relations (24) are necessary and sufficient
  conditions in order that $f^i$  be a generalized-analytical function.
Let us show that these relations can be written in an explicitly invariant form
if the matrix composed of the numbers
\begin{equation}\label{39}
q_{ij}=p^r_{im}p^m_{rj},
\end{equation}
is non-singular, that is if
\begin{equation}\label{40}
q=\det(q_{ij})\ne 0.
\end{equation}
In this case the inverse matrix $(q_{ij})$ forms the tensor
$(q^{ij})$ showing the properties
\begin{equation}\label{41}
q_{jk}q^{ki}= q^{ik}q_{kj}= \delta^i_j.
\end{equation}
Whence, when the formula (22) is applied instead of the formulae
(23) and (24), we get the invariant expression for the derivative
\begin{equation}\label{42}
f'^i=q^{is}p^r_{sm}\tilde{\nabla}_r \cdot f^m
\end{equation}
and for the Cauchy-Riemann relations
\begin{equation}\label{43}
\tilde{\nabla}_kf^i - p^i_{kj}\cdot q^{js}p^r_{sm}
\tilde{\nabla}_rf^m=0.
\end{equation}

    Let us turn to  the generalized-analytical functions $F_{(1)}(X)$ and $F_{(2)}(X)$,
which are constrained by the relation
\begin{equation}\label{44}
F_{(2)}(X)=F(X)\cdot F_{(1)}(X),
\end{equation}
where $F(X)$ - some function of poly-number variable. The function
is generalized-analytical in the field where the function
$F_{(1)}(X)$  is not a divisor of zero. In this case
\begin{equation}\label{45}
F(X)=\frac{F_{(2)}(X)}{F_{(1)}(X)},
\end{equation}
\begin{equation}\label{46}
\tilde{D}F(X)=\frac { F_{(1)}(X) \tilde{D} [F_{(2)}(X)]
 - \tilde{D}[F_{(1)}(X)]F_{(2)}(X)} {F^2_{(1)}(X)}
\end{equation}
or
\begin{equation}\label{47}
F'(X)=\frac{F_{(1)}(X)F'_{(2)}(X)-F'_{(1)}(X)F_{(2)}(X)}
{F^2_{(1)}(X)}.
\end{equation}

    If
\begin{equation}\label{48}
F(X)=F_{(2)}[F_{(1)}(X)],
\end{equation}
then the function $F(X)$  is generalized-analytical with
\begin{equation}\label{49}
F'(X)=F'_{(2)}(F_{(1)})\cdot F'_{(1)}(X).
\end{equation}

\section{Similar geometries and conformal transformations}

Actually, we are interested in not only the pair $\{M_n,P_n\}$ and
generalized-analytical functions $\{f^i,\gamma^i_k\}$ but
(eventually) possible ways of application of these notions to
constructing physical models and solving new physical problems.
Two spaces in which congruences of extremals (geodesics) coincide
are similar in many respects. The extremals are meant to be
solutions to set of equations for definition of curves over which
the length of the curve acquires its extremum; alternatively, they
are meant to be the curves which in a given geometry are defined
to be geodesics (for example, geodesics in geometries of affine
connection). However, for some physical as well as mathematical
problems it is not of great importance which length element is
used in applied space, -- a real use is made to only the set of
equations that define extremals (or to extremals proper). We shall
say that two $n$-dimensional geometries are \textit{similar}, if
there exist such coordinate systems and parameters along curves
that with respect to them the equations for extremals are
equivalent and the initial and/or final date set forth in one
space may also be given in another space.

All the set of generalized-analytic functions can be broken into
the subsets $\{f^i,\Gamma^k_{ij}\}$ that involve the same
connection coefficients $\Gamma^k_{ij}$, so that for all
ge\-ne\-ra\-li\-zed-analytic functions from the subset the
relation
\begin{equation}\label{50}
\Gamma^i_{kj}f^j=\gamma^i_k
\end{equation}
is fulfilled. It should be stressed (once more) that the
coefficients $\Gamma^k_{ij}$ are in\-de\-pen\-dent of any choice
of functions in the subset $\{f^i,\Gamma^k_{ij}\}$. Generally
speaking, the subset may be formed by only one
generalized-analytic function. If $f^i$  and $\gamma^i_k$ are
prescribed, then the relations  (50) can be treated to be a set of
equations for definition of the coefficients $\Gamma^k_{ij}$.
Having find and fixed them, they can be applied for all tensors
and indices, thereafter we get a due possibility to work with the
space of affine connection
 $L_n(\Gamma^k_{ij})$ in which the set of equations for geodesics is of the form
\begin{equation}\label{51}
\frac{d^2x^i}{d \sigma^2}=-\Gamma^i_{kj} \frac{dx^k}{d \sigma}
\frac{dx^j}{d \sigma}.
\end{equation}
Generally speaking, in this way we loose the possibility to use
the poly--number product for construction of new
generalized-analytical functions and should give up the simple
differentiation rules (33). In the last case the covariant
derivative $ \tilde{\nabla}_k $  in the special coordinate system
must act on the tensor $p^i_{kj}$. In order to have simultaneously
on the subset $\{f^i,\Gamma^k_{ij}\}$ the poly-number product of
generalized-analytical functions and the rules (33), which
application yields again a generalized-analytical function, we are
to restrict ourselves to the func\-tions subjected to the
condition (36) with
 $ \Gamma^i_{jk} \equiv {}^{(1)}\Gamma^i_{jk} \equiv {}^{(2)}\Gamma^i_{jk}
 \equiv {}^{(3)}\Gamma^i_{jk}. $

Let us require that the space
 $L_n(\Gamma^i_{jk})$ be similar to a Riemannian or pseudo-Riemannian one
  $V_n(g_{ij})$, where    $g_{ij}$ is a (fundamental) metric tensor. Then instead of  (50)
we get the system of equations
\begin{equation}\label{52}
\left[\frac{1}{2} g^{im}\left( \frac{ \partial g_{km}}{\partial
x^j} +
 \frac{\partial g_{jm}}{\partial x^k} -
 \frac{\partial g_{kj}}{\partial x^m}\right) +
 \frac{1}{2} (p_k \delta^i_j + p_j \delta^i_k) + S^i_{kj} \right] \cdot f^j = \gamma^i_k,
\end{equation}
where  $S^i_{kj}$  stands for an arbitrary tensor (torsion tensor)
obeying the property of skew-symmetry with respect to two indices,
and $p_i$ may be an arbitrary one--covariant tensor \cite{rash1}.
This system may be used to define the fundamental tensor
$g_{ij}$.

There exist such Finslerian spaces which are not of Riemannian or
pseudo-Riemannian type, but in which, however, one has the system
of equations
\begin{equation}\label{53}
\frac{d^2x^i}{d\sigma^2}=
 -\Gamma^i_{kj}[L(dx;x)]\cdot\frac{dx^k}{d\sigma}\frac{dx^j}{d\sigma},
\end{equation}
where the coefficients  $\Gamma^i_{kj}[L(dx;x)]$ are defined by
means of a relevant metric function $L(dx^1,...,dx^n;x^1,...,x^n)$
of Finsler type. The corresponding Finsler spaces are similar to
spaces of affine connection
 endowed with the connection coefficients  $\Gamma^i_{kj}$ deviated possibly from
 the coefficients
$\Gamma^i_{kj}[L(dx;x)]$  by oc\-cur\-ren\-ce of an additive
torsion tensor and/or an additive tensor
$\frac{1}{2}(p_k\delta^i_j+p_j\delta^i_k)$ \cite{rash1}.

Let a  generalized-analytical functions define spaces of the
affine connection $ \, L_n(^{(1)}\Gamma^k_{ij}) \, $ and $ \,
L_n(^{(2)}\Gamma^k_{ij}) \, $ similar
 to corresponding Riemannian or pseudo-Rie\-man\-nian spaces $ \, \, V_n(g_{ij}) \, \, $ and
$\, \, V_n(K^2_Vg_{ij}) \, \, $ \, and/or \, the Finslerian spaces
$ \, \, F_n[L(dx;x)] \, \, $ and $F_n[K_FL(dx,x)]$, where
$K_V(x^1,..,x^n)>0,\,K_F(x^1,...,x^n)>0$ - scalar functions
(invariants). Then the transformation (coordinate and/or in the
space of
 generalized-analytical functions) going over the set  $f^i_{(1)}$  in the set
    $f^i_{(2)}$,
can be called conformal, for under this procedure one has
\begin{equation}\label{54}
g_{ij}(x)\rightarrow K_V^2(x)\cdot  g_{ij}
\end{equation}
and
\begin{equation}\label{55}
L(dx;x)\rightarrow K_F(x)\cdot L(dx;x).
\end{equation}

\section{Possible additional requirements}

From the definition of a generalized-analytical function it
follows that it is possible to present the function by choosing
two arbitrary one--covariant fields $f^i(x^1,...,x^n)$  and
$f'^i(x^1,...,x^n)$. Then the formula (23) entails the following
representation for the gamma-objects:
\begin{equation}\label{56}
\gamma^i_k=-\frac{\partial f^i}{\partial x^k}+p^i_{kj}f'^j .
\end{equation}
The Cauchy-Riemann conditions are fulfilled automatically. So, to
get the field equations for the unknown function-components
$f^i(x^1,...,x^n)$  and $f'^i(x^1,...,x^n)$, it is necessary to
set forth at least $2n$ additional relations, for example, some
partial differential equations of the first--order with respect to
$f^i(x^1,...,x^n)$  and $f'^i(x^1,...,x^n).$

    (1): Let us consider the subset of generalized-analytical functions
$f^i$ such that
\begin{equation}\label{57}
 \tilde{D}F(x)\equiv 0,\,\,\leftrightarrow\,\,\tilde{\nabla}_kf^i\equiv
 0,\,\, \leftrightarrow\,\,f'^i\equiv 0
\end{equation}

In this case the Cauchy-Riemann conditions are fulfilled
automatically and arbitrary vector-function coupled with
 $\gamma^i_k = -\frac{\partial f^i}{\partial x^k}$,
that is the pair $\{f^i,-\frac{\partial f^i}{\partial x^k}\}$, is
a generalized-analytical function. It is important to note that
the properties of poly-numbers do not influence this  procedure.
In other words, this subset (treated on the level of the
Cauchy-Riemann conditions) are independent of any choice of the
system of poly--numbers.

    (2): If instead of the conditions (57) we assume the relations
\begin{equation}\label{58}
\tilde{D}F(X)=\lambda\cdot F(X)\cdot dX, \,\, \leftrightarrow \,\,
\tilde{\nabla}_kf^i = \lambda\cdot p^i_{kj}\cdot f^j, \,\,
\leftrightarrow \,\, f'^i=\lambda\cdot f^i,
\end{equation}
where $\lambda$ is a real number, then the pairs
$\{f^i,-\frac{\partial f^i}{\partial x^k} + \lambda p^i_{kj}f^j\}$
with arbitrary vector-functions
 $f^i$  will form the subset of
the  generalized-analytical functions which to some extent account
for properties of poly-numbers.

    (3):   Farther generalizing the requirements (57) and (58) can be formulated in
the form
\begin{equation}\label{59}
F'(X)=\Lambda\cdot F(X),
\end{equation}
where
\begin{equation}\label{60}
\Lambda = \lambda^1 e_1 + \lambda^2 e_2 +...+ \lambda^n e_n
\end{equation}
an arbitrary poly-number. In this case the pair
\begin{equation}\label{61}
\left\{f^i,-\frac{\partial f^i}{\partial x^k} +
p^i_{kj}p^j_{mr}\lambda^m f^j\right\}
\end{equation}
will be the generalized-analytical functions.

    (4): Using the formulas (23) and (24), we can prove the following statement. If the
relations
\begin{equation}\label{62}
1) \qquad \Gamma^i_{kj}f^j=\gamma^i_k, \qquad \qquad \qquad \qquad
\qquad \qquad \qquad \qquad \qquad \qquad \qquad \qquad \qquad
\qquad
\end{equation}
\begin{equation}\label{63}
2) \qquad \Gamma^i_{1j}p^j_{kr}-p^i_{kj}\Gamma^j_{1r}=0,  \qquad
\qquad \qquad \qquad \qquad \qquad \qquad \qquad \qquad \qquad
\qquad \qquad
\end{equation}
\begin{equation}\label{64}
3) \qquad \frac{\partial \Gamma^i_{1r}}{\partial x^k} -
\frac{\partial \Gamma^i_{kr}}{\partial x^1} +
\left[(\Gamma^i_{kj}-p^i_{km}\Gamma^m_{1j})
 \Gamma^j_{1r} - \Gamma^i_{1j}(\Gamma^j_{kr}  - p^j_{km}\Gamma^m_{1r}) \right] = 0
\end{equation}
hold, then together with the  generalized-analytical pair
$\{f^i,\gamma^i_k\},$ the pair
\begin{equation}\label{65}
 \{f'^i,\Gamma^i_{kj}f'^j\},\{f''^i,\Gamma^i_{kj}f''^j\},...,\{f^{(m)i},
 \Gamma^i_{kj}f^{(m)j}\}, ...
\end{equation}
are also generalized-analytical. In the last formulas the notation
\begin{equation}\label{66}
f^{(m)i}\equiv\frac{\partial f^{(m-1)i}}{\partial x^1} +
\Gamma^i_{1j}f^{(m-1)j}
\end{equation}
has been used.

    (5):   One additional requirements can sound: for the subset
 $\{f^i, \Gamma^i_{kj}\}$
 of generalized-analytical functions a Riemannian or pseudo-Riemannian geometry $V_n(g_{ij})$
similar to the affine connection geometry $L_n(\Gamma^i_{jk})$ can
be found.

    (6):   If a Finsler space $F_n[L(dx;x)]$ is similar to a space of
affine connection, then one among possible requirements can claim
that the subset $\{f^i,\Gamma^i_{jk}\}$ give rise to an affine
connection geometry similar to the Finsler geometry
$F_n[L(dx;x)]$.

    (7):    Let
\begin{equation}\label{67}
x^i=x^i(\tau)
\end{equation}
be a parametric presentation of some curve joining two points
$x^i_{(0)}=x^i(0),\, x^i_{(1)}=x^i(1),$ that is, the parameter
along curves varies in the limits $\tau\in [0;1]$. Let us consider
the functional with integration along indicated curve
\begin{equation}\label{68}
\begin{array}{l}
I[x^i(\tau)] = \displaystyle\int^1_0F(X) \,dX =
\left[\int^1_0p^i_{kj}f^k(x^1(\tau),...,x^n(\tau))dx^j\right]\cdot e_i= \\
 \\
\qquad \qquad            =\left[\displaystyle\int^1_0
p^i_{kj}f^k\frac{dx^j}{d\tau}\right]\cdot e_i,
\end{array}
\end{equation}
where $F(X)$ - some generalized-analytical function, and require
that
 value of the integral (68) be
 independent of integration way, in which case
 the variation of this functional at fixed ends
 of curves should vanish,
 that is the Euler conditions
\begin{equation}\label{69}
\frac{d}{d\tau}\left(p^i_{kj}f^j\right)-p^i_{mj} \frac{\partial
 f^j}{\partial x^k}\frac{dx^m}{d\tau}=0
\end{equation}
or
\begin{equation}\label{70}
\left(p^i_{kj}\frac{\partial f^j}{\partial x^m} - p^i_{mj}
\frac{\partial f^j}{\partial x^k} \right) \cdot
\frac{dx^m}{d\tau}=0
\end{equation}
must be valid. Assuming that $x^i(\tau)$  are arbitrary smooth
functions, from these equations we get
\begin{equation}\label{71}
p^i_{kj}\frac{\partial f^j}{\partial  x^m} -
p^i_{mj}\frac{\partial f^j}{\partial x^k}=0,
\end{equation}
or, recollecting that $\{f^i,\gamma^i_k\}$  is a
generalized-analytic pair,
\begin{equation}\label{72}
p^i_{kj}\gamma^i_m - p^i_{mj}\gamma^j_k =0.
\end{equation}
From these relations it ensues that for the functions $f^i$ the
Cauchy-Riemann conditions (11) hold fine.

Thus, the assumption of independence of the integral (68) of the
path leads to the conclusion that the function $F(X)$  is
analytical, that is such an assumption is superfluous for
non-trivial generalization of the concept of analyticity.

\section{Case $H_4$}

    It is convenient to work with the associative-commutative hypercomplex numbers
in term of the $\psi$-basis which relates to the basis
\begin{equation}\label{73}
e_1=1, \,e_2=j, \,e_3=k, \,e_4=jk, \qquad j^2=k^2=(jk)^2=1
\end{equation}
by means of the linear dependence
\begin{equation}\label{74}
e_i=s^j_i\cdot\psi_j,
\end{equation}
where
\begin{equation}\label{75}
s^j_i=\left(
 \begin{array}{rrrr}
 1&1&1&1 \\
 1&1&-1&-1 \\
 1&-1&1&-1 \\
 1&-1&-1&1 \\\end{array} \right) , \qquad s^k_i\cdot
 s^j_k=4\cdot\delta^j_i.
\end{equation}

    For the basis elements $ \psi_1, \,\psi_2, \,\psi_3, \,\psi_4 $ the
multiplication law
\begin{equation}\label{76}
\psi_i\cdot\psi_j=p_{ij}^{(\psi)k}\cdot\psi_k
\end{equation}
involves the characteristic numbers
\begin{equation}\label{77}
p_{ij}^{(\psi)k}=\left\{
 \begin{array}{ll}
 1,&\;\mbox{if}\;i=j=k, \\
 0,&\;\mbox{in other cases}.
 \end{array}\right.
\end{equation}
We shall use the following notation:
\begin{equation}\label{78}
X=x^1e_1+...+x^4e_4=\xi^1\psi_1+...+\xi^4\psi_4
\end{equation}
and
\begin{equation}\label{79}
F(X)=\varphi^1(\xi^1,...,\xi^4)\cdot\psi_1+...+\varphi^4(\xi^1,...,\xi^4)\cdot\psi_4.
\end{equation}

   Thus, if $\, \varphi^i(\xi^1,...,\xi^4) $ -
a generalized-analytical function of the  $H_4$-variable used, then
such a vector-function
$\varphi'^i(\xi^1,...,\xi^4) $ can be found that
\begin{equation}\label{80}
\frac{\partial\varphi^i}{\partial\xi^k}+\gamma_k^{(\psi)i}=p_{kj}^{(\psi)i}\cdot\varphi'^j.
\end{equation}

    Taking into account (77), we get
\begin{equation}\label{81}
\left.
\begin{array}{lll}
\displaystyle\frac{\partial\varphi^1}{\partial\xi^1}+\gamma_1^{(\psi)1}=\varphi'^1,&
\displaystyle\frac{\partial\varphi^1}{\partial\xi^2}+\gamma_2^{(\psi)1}=0,&
\displaystyle\frac{\partial\varphi^1}{\partial\xi^3}+\gamma_3^{(\psi)1}=0, \\
\displaystyle\frac{\partial\varphi^1}{\partial\xi^4}+\gamma_4^{(\psi)1}=0, & & \\
\displaystyle\frac{\partial\varphi^2}{\partial\xi^1}+\gamma_1^{(\psi)2}=0,&
\displaystyle\frac{\partial\varphi^2}{\partial\xi^2}+\gamma_2^{(\psi)2}=\varphi'^2,&
\displaystyle\frac{\partial\varphi^2}{\partial\xi^3}+\gamma_3^{(\psi)2}=0, \\
\displaystyle\frac{\partial\varphi^2}{\partial\xi^4}+\gamma_4^{(\psi)2}=0, & &  \\
\displaystyle\frac{\partial\varphi^3}{\partial\xi^1}+\gamma_1^{(\psi)3}=0,&
\displaystyle\frac{\partial\varphi^3}{\partial\xi^2}+\gamma_2^{(\psi)3}=0,&
\displaystyle\frac{\partial\varphi^3}{\partial\xi^3}+\gamma_3^{(\psi)3}=\varphi'^3, \\
\displaystyle\frac{\partial\varphi^3}{\partial\xi^4}+\gamma_4^{(\psi)3}=0, & & \\
\displaystyle\frac{\partial\varphi^4}{\partial\xi^1}+\gamma_1^{(\psi)4}=0,&
\displaystyle\frac{\partial\varphi^4}{\partial\xi^2}+\gamma_2^{(\psi)4}=0,&
\displaystyle\frac{\partial\varphi^4}{\partial\xi^3}+\gamma_3^{(\psi)4}=0, \\
\displaystyle\frac{\partial\varphi^4}{\partial\xi^4}+\gamma_4^{(\psi)4}=\varphi'^4.
& &
 \end{array}\right\}
\end{equation}

These relations involve the expression for the derivative
\begin{equation}\label{82}
\varphi'^i=\frac{\partial\varphi^i}{\partial\xi^{i_{-}}}+\gamma_{i_{-}}^{(\psi)i}
\end{equation}
($i=i_-$, for which  no summation is assumed), and also the
Cauchy-Riemann relations
\begin{equation}\label{83}
\frac{\partial\varphi^i}{\partial\xi^k} +
\gamma_k^{(\psi)i}=0,\;i\neq k.
\end{equation}

    The space $H_4$ is the metric (Finslerian) space in which the length element
$ds$  is expressible through the form $d\xi^1d\xi^2d\xi^3d\xi^4$   in a conic
region defined possibly in various ways. Let us stipulate that
\begin{equation}\label{84}
ds=\sqrt[4]{d\xi^1d\xi^2d\xi^3d\xi^4},
\end{equation}
assuming that the region is prescribed by the inequalities
\begin{equation}\label{85}
d\xi^1\geq0,\;d\xi^2\geq0,\;d\xi^3\geq0,\;d\xi^4\geq0\;.
\end{equation}

    Let us consider the four-dimensional Finslerian geometry with the
length element of the form
\begin{equation}\label{86}
ds=\sqrt[4]{\kappa^4\cdot d\xi^1d\xi^2d\xi^3d\xi^4},
\end{equation}
where $\kappa \equiv \kappa(d\xi^1d\xi^2d\xi^3d\xi^4)>0. $ Such a
geometry is not Riemannian or pseudo-Riemannian. Let us show that
such a geometry is similar (according to terminology adopted
above) to some affine geometry with a connection
$L_4(\Gamma_{kj}^i).$ Let us write equations for extremals of this
Finslerian space by using the tangential equation of indicatrix
\cite{rash2}:
\begin{equation}\label{87}
\Phi(p_1,...,p_4;\xi^1,...,\xi^4)=0\;,
\end{equation}
where
\begin{equation}\label{88}
Phi(p;\xi)=p_1p_2p_3p_4-\left(\frac{\kappa}{4}\right)^4\;,
\end{equation}
and
\begin{equation}\label{89}
p_i=\frac{\partial(ds)}{\partial(d\xi^i)}=\frac{1}{4}\cdot\frac{\sqrt[4]{\kappa^4\cdot
d\xi^1d\xi^2d\xi^3d\xi^4}}{d\xi^i}.
\end{equation}
Then the set of equations for definition of extremals reads
\begin{equation}\label{90}
 \left.
\begin{array}{c}
\displaystyle\frac{d\xi^1}{\displaystyle\frac{\partial\Phi}{\partial
p_1}}=...=\displaystyle\frac{
d\xi^4}{\displaystyle\frac{\partial\Phi}{\partial p_4}}=
\displaystyle\frac{
dp_1}{-\displaystyle\frac{\partial\Phi}{\partial \xi^1}}=...=
\displaystyle\frac{dp_4}{-\displaystyle\frac{\partial\Phi}{\partial \xi^4}}, \\
\\
\Phi(p,\xi)=0;
\end{array}\right\}
\end{equation}
or
\begin{equation}\label{91}
d\xi^i=\frac{\partial \Phi}{\partial p_i}\cdot\lambda\cdot
d\tau,\; dp_i=-\frac{\partial \Phi}{\partial
\xi^i}\cdot\lambda\cdot d\tau,\;\;\Phi(p;\xi)=0,
\end{equation}
where $\tau$ - a parameter along extremals, and
$\lambda\equiv\lambda(p;\xi)\ne 0$  -- a function. For the
tangential equation of the indicatrix (87), (88) the set of
equations (91) takes on the form
\begin{equation}\label{92}
\dot{\xi}^i=\frac{p_1p_2p_3p_4}{p_i}\cdot\lambda, \quad
 \dot{p}^i=\frac{1}{4^4}\frac{\partial k^4}{\partial \xi^i}\cdot\lambda, \quad
 p_1p_2p_3p_4=\left( \frac{k}{4} \right)^4,
\end{equation}
with
\begin{equation}\label{93}
\dot{\xi}^i=\frac{d\xi^i}{d\tau},\;\;
\dot{p}_i=\frac{dp_i}{d\tau}\, .
\end{equation}
Let us consider $\lambda=\lambda(\xi)>0$ to be a function of only
coordinates. Then, by explicating $p_i$, we get the set of
equations for definition of extremals in the Finslerian space (86)
in the form
\begin{equation}\label{94}
\ddot{\xi}^i = -\Gamma^i_{kj} \dot{\xi}^k \dot{\xi}^j,
\end{equation}
where
\begin{equation}\label{95}
\Gamma^i_{kj}=-\left\{
\begin{array}{ll}
\displaystyle\frac{\partial ln\left(\displaystyle\frac{\lambda}
{\lambda_0}\right)}{\partial\xi^j},&\mbox{if}\;i=j=k, \\
\delta^i_k \displaystyle\frac{\partial ln\left(\displaystyle
\frac{\sigma}{\sigma_0}\right)}{\partial\xi^j},&\mbox{in other
cases;}
\end{array}
\right.
\end{equation}
\begin{equation}\label{96}
\sigma=\left(\frac{\kappa}{4}\right)^4\cdot\lambda,
\end{equation}
$\lambda_0$ and $\sigma_0$ are constants of relevant dimensions.
Let us write down explicitly the coefficients $\Gamma^i_{kj}$:
\begin{equation}\label{97}
(\Gamma^1_{kj})=-\left(
\begin{array}{cccc}
\displaystyle\frac{\partial
ln\left(\frac{\lambda}{\lambda_0}\right)}{\partial\xi^1}&
\displaystyle\frac{\partial
ln\left(\frac{\sigma}{\sigma_0}\right)}{\partial\xi^2}&
\displaystyle\frac{\partial
ln\left(\frac{\sigma}{\sigma_0}\right)}{\partial\xi^3}&
\displaystyle\frac{\partial ln\left(\frac{\sigma}{\sigma_0}\right)}{\partial\xi^4} \\
0&0&0&0 \\
0&0&0&0 \\
0&0&0&0
\end{array}
\right),
\end{equation}
\begin{equation}\label{98}
(\Gamma^2_{kj})=-\left(
\begin{array}{cccc}
0&0&0&0 \\
\displaystyle\frac{\partial
ln\left(\frac{\sigma}{\sigma_0}\right)}{\partial\xi^1}&
\displaystyle\frac{\partial
ln\left(\frac{\lambda}{\lambda_0}\right)}{\partial\xi^2}&
\displaystyle\frac{\partial
ln\left(\frac{\sigma}{\sigma_0}\right)}{\partial\xi^3}&
\displaystyle\frac{\partial ln\left(\frac{\sigma}{\sigma_0}\right)}{\partial\xi^4} \\
0&0&0&0 \\
0&0&0&0
\end{array}
\right),
\end{equation}
\begin{equation}\label{99}
(\Gamma^3_{kj})=-\left(
\begin{array}{cccc}
0&0&0&0 \\
0&0&0&0 \\
\displaystyle\frac{\partial
ln\left(\frac{\sigma}{\sigma_0}\right)}{\partial\xi^1}&
\displaystyle\frac{\partial
ln\left(\frac{\sigma}{\sigma_0}\right)}{\partial\xi^2}&
\displaystyle\frac{\partial
ln\left(\frac{\lambda}{\lambda_0}\right)}{\partial\xi^3}&
\displaystyle\frac{\partial ln\left(\frac{\sigma}{\sigma_0}\right)}{\partial\xi^4} \\
0&0&0&0
\end{array}
\right),
\end{equation}
\begin{equation}\label{100}
(\Gamma^4_{kj})=-\left(
\begin{array}{cccc}
0&0&0&0 \\
0&0&0&0 \\
0&0&0&0 \\
\displaystyle\frac{\partial
ln\left(\frac{\sigma}{\sigma_0}\right)}{\partial\xi^1}&
\displaystyle\frac{\partial
ln\left(\frac{\sigma}{\sigma_0}\right)}{\partial\xi^2}&
\displaystyle\frac{\partial
ln\left(\frac{\sigma}{\sigma_0}\right)}{\partial\xi^3}&
\displaystyle\frac{\partial
ln\left(\frac{\lambda}{\lambda_0}\right)}{\partial\xi^4}
\end{array}
\right).
\end{equation}

    It will be noted that instead of the matrices (97) - (100) one can take
their transforms. Thus, the Finslerian geometry with the length
element (86) is similar to the geometry of the affine connection
$L_4[\Gamma^i_{kj}+S^i_{kj}+\frac{1}{2}(p_k\delta^i_j+p_j\delta^i_k)],$
where  $S^i_{kj}$ --  a tensor which is assumed to be
skew-symmetric with respect to the subscripts, and  $p_k$ stands
for an arbitrary one-covariant tensor.

Let us consider the generalized-analytical functions $\varphi^i$
of $H_4$-variable that obey the additional condition  3), that is
the pair
\begin{equation}\label{101}
\left\{\varphi^i,-\frac{\partial\varphi^i}
{\partial\xi^k}+p_{kj}^{(\psi)i}\mu^j\varphi^j\right\},
\end{equation}
where
\begin{equation}\label{102}
\Lambda=\lambda^i\cdot e_i=\mu^j\cdot\psi_j.
\end{equation}
Let us select from such pairs a subset
$\{\varphi^i,\Gamma^i_{kj}\}$, where $\Gamma^i_{kj}$ are defined
by the matrices transposed to the matrices (97) - (100). In this
way, the requirement 6) is retained. Then the pair (101) should
fulfill the 16 relations (50) the first four of which are
\begin{equation}\label{103}
\left.
\begin{array}{ll}
  \displaystyle\frac{\partial\varphi^1}{\partial\xi^1}=\mu^1\varphi^1 + \frac{
 \partial\, ln\left(\frac{\lambda}{\lambda_0}\right)}{\partial
 \xi^1}\varphi^1, &
 \displaystyle\frac{\partial\varphi^1}{\partial\xi^2}=
 \frac{\partial\, ln\left(\frac{\sigma}{\sigma_0}\right)}{\partial \xi^2}\varphi^1,
 \\
 \qquad & \qquad  \\
  \displaystyle\frac{\partial\varphi^1}{\partial\xi^3}=\frac{\partial\,
 ln\left(\frac{\sigma}{\sigma_0}\right)}{\partial \xi^3}\varphi^1,
 &
  \displaystyle\frac{\partial\varphi^1}{\partial\xi^4}=\frac{\partial\,
 ln\left(\frac{\sigma}{\sigma_0}\right)}{\partial \xi^4}\varphi^1.
\end{array}
\right\}
\end{equation}

    For the compatibility it is necessary and sufficient that the mixed derivatives
obtained with the help of the formulae  (103) be equal. A part of
these equations, except for three ones, is automatically
satisfied. If we consider all the 16 equa\-tions, not confining
ourselves to the first four equations, we get the following 12
conditions:
\begin{equation}\label{104}
\frac{\partial^2
 ln\left(\frac{\kappa}{\kappa_0}\right)^4}{\partial \xi^i \partial \xi^j}=0, \quad i \neq j;
\end{equation}
from which it ensues that
\begin{equation}\label{105}
ln\left(\frac{\kappa}{\kappa_0}\right)^4=a_1(\xi^1)+a_2(\xi^2)+a_3(\xi^3)+a_4(\xi^4)
\end{equation}
or
\begin{equation}\label{106}
\kappa=\kappa_0\cdot
\exp\{[a_1(\xi^1)+a_2(\xi^2)+a_3(\xi^3)+a_4(\xi^4)]/4\},
\end{equation}
where $a_i$ are four arbitrary functions of one real argument.
Then from equations  (103) and relevant equations for other
components of the generalized-analytical function, we get
\begin{equation}\label{107}
\varphi^i=\varphi^i_{(0)}\left(\frac{\kappa}
{\kappa_0}\right)^4\left(\frac{\lambda}{\lambda_0}\right)
          b_{i_-}(\xi^{i_-})\cdot exp(\mu^{i_-}\xi^{i_-}),
\end{equation}
where
\begin{equation}\label{108}
a_i(\xi^{i_-})=ln\left|b_i(\xi^{i_-})\right|.
\end{equation}

    Thus, despite of two additional requirement,
the generalized-analytical func\-tion (107) in general case is not
reducible to an analytical function of $H_4$-variable, and besides
we obtain the expression (106) for the coefficients $\kappa$ in
the metric function of the Finslerian space with the length
element (86). If
\begin{equation}\label{109}
\frac{\lambda}{\lambda_0}=\left(\frac{\kappa_0}{\kappa}\right)^4,
\end{equation}
then $\varphi^i$ is an analytical function.

    If
\begin{equation}\label{110}
\kappa=\kappa_0\cdot
 \exp\{[(\xi^1)^2+(\xi^2)^2+(\xi^3)^2+(\xi^4)^2]/4\},
\end{equation}
then with respect to the coordinates $x^i$
\begin{equation}\label{111}
\kappa=\kappa_0\cdot exp\{(x^1)^2+(x^2)^2+(x^3)^2+(x^4)^2\}.
\end{equation}

\section{Conclusion}

Having introduced the concept of the generalized-analytical
function of poly-number variable in the present work, we have made
the first step in the direction of constructing a relevant theory
aiming to develop theoretical-physical models. An important
ingredient of such investigations must be search for additional
requirements to be obeyed by the generalized-analytical functions
and for the consequences implied by the requirements. The
conditions that lead to trivial results --- to analytical
functions --- should especially be analyzed. This may admit
formulating the properties that are forbidden to  attribute proper
generalized-analytical functions of poly-number variable (in
contrast to  analytical functions proper). As it has been shown
above, the independence of integral of integration path relates to
such properties. Of course, it is necessary to carry out a
particular attentive study to compare the properties of analytical
functions of complex variable and generalized-analytical functions
of poly-number variable in case of the dimension exceeding 2. It
can be hoped, therefore, that the concepts and results of the
present work may face future novel theoretical-physical
applications.

\end{document}